\documentclass[12pt,prl]{revtex4}

\usepackage{bm}                         
\usepackage{graphicx}

\newcommand{\PRG}{PuRhGa$_5$}
\newcommand{\PCG}{PuCoGa$_5$}

\begin{document}

\title{Theory of Knight Shift and Spin-Lattice Relaxation Rates of Pu-115}
\author{Yunkyu Bang}
\affiliation{Department of Physics, Chonnam National University, Kwangju 500-757, Korea}

\author{Matthias J. Graf, Nicholas J. Curro, and Alexander V. Balatsky}
\affiliation{Los Alamos National Laboratory, Los Alamos, New Mexico 87545, USA}

\maketitle

\section{Abstract}

We calculated the Knight shift and spin-lattice relaxation rates of Pu-115 compounds 
assuming d-wave superconductivity in the presence of strong impurity scattering.  We discuss 
the implications for recent measurements of the spin-lattice relaxation rate in the Pu-115
compound \PRG\ by Sakai and coworkers [J. Phys. Soc. Jpn. {\bf 74}, 1710 (2005)]
and present a prediction for the corresponding Knight shift. In addition, we
noticed a significant round-off of the spin-lattice relaxation rate 1/T$_1$ just above the
superconducting transition temperature that is not observed in the sister compound \PCG.
It appears that in \PRG\ superconductivity is mediated by spin fluctuations, too. 
This provides additional support to the scenario of superconducting pairing mediated by spin 
fluctuations in the Pu-115 compounds similar to the Ce-115 compounds and the
high-temperature copper-oxide superconductors.

\section{Introduction}

The discovery of superconductivity in plutonium based systems such as
\PCG \cite{Sarrao} and \PRG \cite{Wastin} has stimulated the study of unconventional 
superconductivity and the pairing symmetry and mechanism in these materials.
The symmetry of an unconventional superconductor is reduced compared to the symmetry
of its normal state, thus resulting in many novel properties of the quasiparticle excitation spectrum.
It is believed that the superconducting action in Pu-115 [Pu{\it M}Ga$_5$ with {\it M}=Co and Rh]
derives itself from the unique character of the 5f electrons of plutonium \cite{Bang04a}.
The tetragonal crystal structure of Pu{\it M}Ga$_5$ is isostructural to that of
the Ce-115 series [Ce{\it M}In$_5$]. 

The purpose of this study is to shed light on the superconducting pairing symmetry and 
possible pairing mechanism in the Pu-115 compounds.
Very recently,
Curro and coworkers \cite{Curro} proposed, based on their measurements of the Knight shift
and spin-lattice relaxation rates, that the Pu-115 compounds are bridging the 
superconducting and normal-state properties of the heavy-fermion Ce-115 and high-temperature
copper-oxide superconductors. Therefore providing a means for tuning the interaction
strength of antiferromagnetic spin fluctuations to intermediate values between both extreme
limits \cite{Bauer04}.

The experimental techniques of nuclear magnetic resonance (NMR) and nuclear quadrupolar 
resonance (NQR) have been used successfully in the past to distinguish between the 
spin states of Cooper pairs (spin singlet vs. spin triplet pairing) and provide indirect
information on the symmetry of the gap function -- fully gapped vs.  nodal lines or nodal
points in the gap function on the Fermi surface. 
Both techniques probe directly the quasiparticle density of states and reveal indirect 
information about the pairing symmetry.

The standard explanation of power vs. exponential laws in the low-temperature behavior
of thermodynamic and transport properties, for example, the spin-lattice relaxation rate
$1/T_1$, comes from the difference of nodal and fully gapped excitation spectra in the
superconducting state.

In clean nodal superconductors $1/T_1$ exhibits a nearly $T^3$ behavior  
far below the superconducting transition temperature $T_{c}$, while it is exponential for
gapped superconductors. 
On the other side,  deviations from this behavior, like the $T$-linear temperature dependence 
of $1/T_1$ at low temperatures, are explained by impurity effects in an 
unconventional superconductor (SC) with lines of nodes on the Fermi surface.
Very recently, Sakai and coworkers \cite{Sakai} reported
such a result for the spin-lattice relaxation rate of \PRG\ in the superconducting
phase. This behavior closely resembles the spin-lattice relaxation rates measured by
Curro and coworkers for \PCG, which belongs to the same family of Pu-115 compounds,
albeit with a $T_c$ nearly twice as high \cite{Curro}.

Here we give a detailed theoretical description of the spin-lattice relaxation rate
and predict what should be observed for the Knight shift if measured on the same sample.
Our self-consistent treatment of impurity scattering in the superconducting state
goes beyond the two-fluid approach used by Sakai et al. \cite{Sakai}, which was used to
explain the large residual density of states in \PRG.
Simultaneous measurements of spin-lattice relaxation rate and Knight shift will place
stringent constraints on the symmetry and magnitude of the superconducting gap function,
as well as on the concentration and scattering strength of impurities at low temperatures.

\section{Theory}

The effect of impurity scattering is included within the self-consistent $T$-matrix
approximation \cite{Bang03,Bang04}, which is the standard formulation for pointlike 
defects in a superconducting dilute alloy \cite{Hirschfeld86,SchmittRink86,Monien87}. 
For the case of particle-hole symmetry of the quasiparticle excitation spectrum 
the Nambu component $T_3$ of the $T$ matrix vanishes, 
and for a d-wave order parameter (OP) with isotropic scattering $T_1=0$
(also without loss of generality we can choose $T_2=0$ by general U(1) gauge
symmetry), where $T_i$ is the $i$th component of the $2\times 2$
Nambu matrix expanded in Pauli matrices. Then we need to calculate only
$T_0(\omega)$. The impurity self-energy is given by
$\Sigma_{0}=\Gamma T_{0}$, where $\Gamma=n_i/\pi N_{0}$. Here $N_0$ is
the normal density of states (DOS) at the Fermi surface (FS), $n_i$ is the impurity
concentration;  $T_0 (\omega_n) =\frac{g_0 (\omega_n)}{[c^2-g_0 ^2
(\omega_n)]}$, where $g_0 (\omega_n) = \frac{1}{\pi N_0}  \sum_k
\frac{i \tilde{\omega}_n}{\tilde{\omega}_n^2 + \epsilon_k^2
+\Delta^2(k)}$. 
The impurity renormalized Matsubara frequency is defined by
 $\tilde{\omega}_n=\omega_n+\Sigma_0$, with $\omega_n=\pi T (2n+1)$, 
and the scattering strength parameter $c$ is related to the s-wave phase 
shift $\delta_0$  by $c=\cot(\delta_0)$.
Using this self-energy $\Sigma_0$ the following gap equation is solved self-consistently,
\begin{eqnarray}
\label{eq_gap}
\Delta(\phi) &=& - N_0 g(\phi) \int \frac{d \phi^{'}}{2 \pi}
V(\phi-\phi^{'})  
  T \sum_{\omega_n} \int^{\omega_c}_{-\omega_c} d \epsilon
  \frac{
    \Delta(\phi^{'})}{\tilde{\omega}_n^2 + \epsilon^2
    +\Delta^2(\phi^{'})}  
\ ,
\end{eqnarray}
where $V(\phi-\phi')$ is the angular parametrization of the
pairing interaction, and $\omega_c$ is a typical cutoff energy.
We assume the canonical d-wave gap function of the form 
$\Delta(\vec{k})= \Delta_0 (\cos k_x -\cos k_y)$ or 
$\Delta(\phi) = \Delta_0 \cos(2 \phi)$ for a cylindrical Fermi surface.
The pairing potential $V(\phi-\phi^{'})$  induces a gap with d-wave symmetry. 
Although its microscopic origin is not the issue of this paper, we
believe it originates from  antiferromagnetic (AFM) spin fluctuations. 
The static limit of the spin susceptibility of the AFM fluctuations, 
$\chi({\bf q},\omega=0) \sim \frac{1}{(\bm{q}-\bm{Q})^2+\xi^{-2}}$, 
is parameterized near the AFM wave vector $\bm{Q}$ as \cite{Bang}
\begin{equation}
\label{eq_pairing}
V(\phi-\phi^{'})=V_d(b) \frac{b^2}{(\phi-\phi^{'} \pm \pi/2)^2+b^2}
\ ,
\end{equation}
where the parameter $b$ is inverse proportional to the AFM correlation 
length $\xi$, normalized by the cylindrical FS $(\xi \sim a \pi/ b$; $a$ is the
lattice parameter). For all calculations in this paper, we chose
$b=0.5$ which is not a sensitive parameter for our results unless
$\xi$ is very large ($ b < 0.1$) \cite{Bang}, i.e.,  within the
range of $0.1 < b < 1$ our results show little variations and
are qualitatively the same.

With the gap function $\Delta(\phi)$ and  $T_0 (\omega)$ obtained from Eq.~(\ref{eq_gap})
($T_0 (\omega)$ is analytically continued from $T_0 (\omega_n)$ by Pad\'e
approximant method), we calculate the $1/T_1$ nuclear spin-lattice relaxation
rate  \cite{Hirschfeld88,Bang03,Bang04,Curro}
\begin{eqnarray}
\label{eq_spin_lattice_rate}
\frac{1}{T_1 T} &\sim&  -\int_0 ^{\infty} \frac{\partial f_{F}
(\omega)}{\partial \omega}
\left[
 \left\langle Re
   \frac{\tilde{\omega}}{\sqrt{\tilde{\omega}^2-\Delta^2(\phi)}}
  \right\rangle_{\phi}^2
 + \left\langle Re
     \frac{\Delta(\phi)}{\sqrt{\tilde{\omega}^2-\Delta^2(\phi)}}
    \right\rangle_{\phi}^2
\right],
\end{eqnarray}
and the superconducting spin susceptibility $\chi_S$
\begin{eqnarray}
\label{eq_spin_susceptibility}
\frac{\chi_S}{T} &\sim&  -\int_0 ^{\infty} \frac{\partial f_{F}
(\omega)}{\partial \omega}
 \left\langle Re
   \frac{\tilde{\omega}}{\sqrt{\tilde{\omega}^2-\Delta^2(\phi)}}
  \right\rangle_{\phi}
,
\end{eqnarray}
where $f_F(\omega)$ is the Fermi-Dirac function, the impurity renormalized
quasiparticle energy
$\tilde{\omega}=\omega+\Sigma_0(\omega)$, and
$\langle...\rangle_{\phi}$ means the angular average over the FS. The first
term in the bracket of Eq.~(\ref{eq_spin_lattice_rate}) 
is  $N^2 (\omega)$. The second term vanishes in
our calculations because of the symmetry of the OP.  To calculate $1/T_1 T$
using Eq.~(\ref{eq_spin_lattice_rate}), or $\chi_S$ using Eq.~(\ref{eq_spin_susceptibility}),
we need the full temperature dependent gap function
$\Delta(\phi,T)$ and $T_{c}$. Our gap equation Eq.~(\ref{eq_gap}) is the BCS
gap equation, therefore it gives  the BCS temperature behavior for
$\Delta(\phi,T)$ and $\Delta_0=2.14 \, {k_B T_c}$ for the standard
weak-coupling d-wave SC. In order to account for strong-coupling effects 
we use the phenomenological formula
$\Delta(\phi,T)=\Delta(\phi,T=0)~ \Xi(T)$ with $\Xi(T)=\tanh (\beta
\sqrt{T_{c}/T-1})$, and parameters $\beta$ and $\Delta_0/ T_{c}$.   Then we
only need to calculate $\Delta(\phi,0)$ at zero temperature. The temperature
dependence of $\Sigma_0(\omega,T)$ ($=\Gamma~ T_0 (\omega,T)$) is similarly
extrapolated: $T_0(\omega,T)=T_0(\omega,T=0)~ \Xi(T) + T_{normal}(1-\Xi(T))$,
where $T_{normal}=\Gamma/(c^2+1)$ is the normal state $T_0$. In our numerical
calculations we chose $\beta=1.74$, because our final results are not very
sensitive with respect to this parameter,
while the ratio $\Delta_0/ k_B T_{c}$ is an important parameter
to simulate strong-coupling effects.  The larger this ratio is, the more 
important are strong-coupling effects.

\section{Results and Discussions}
\begin{figure}
\centerline{ \includegraphics[height=3.5in]{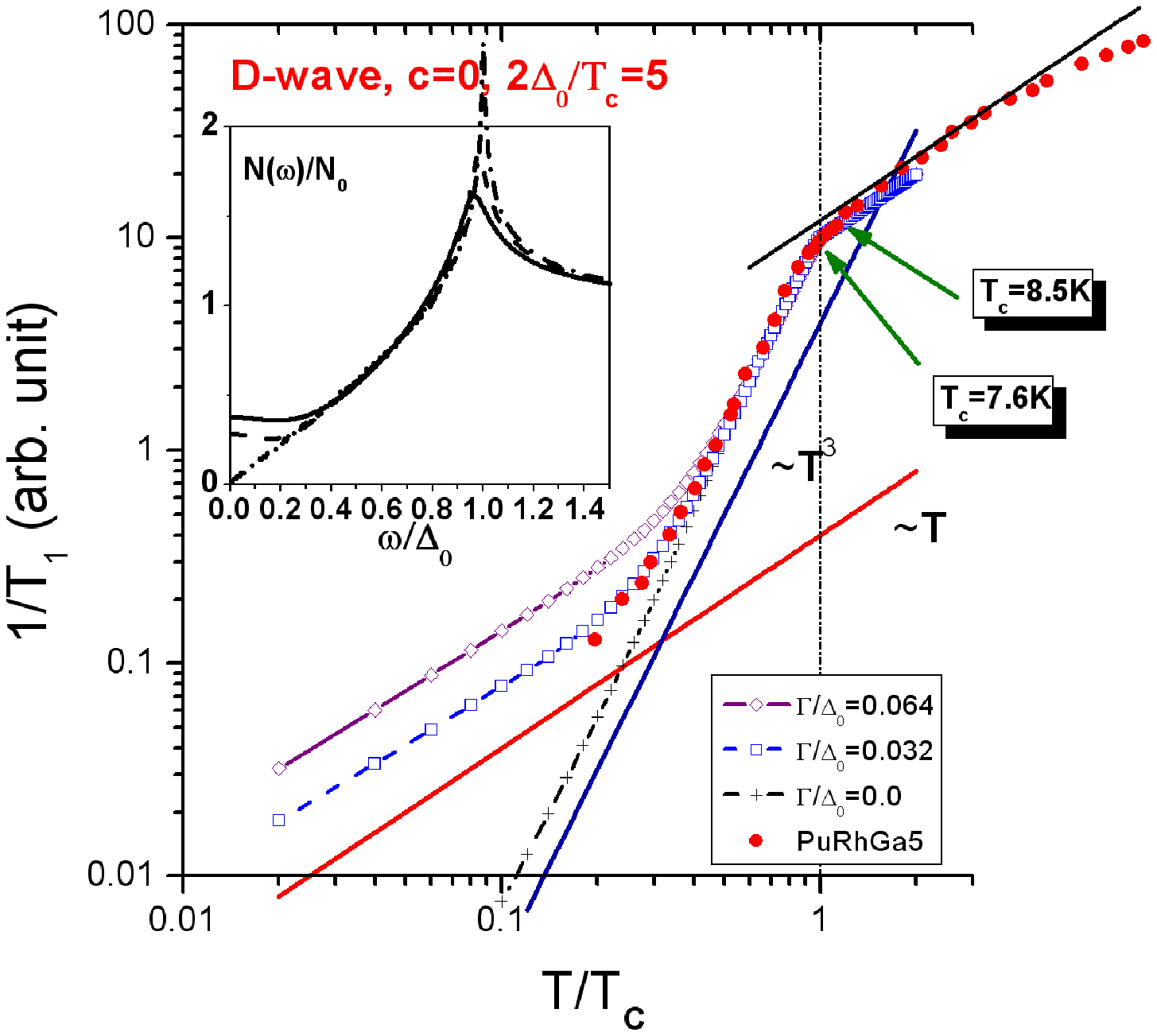} }
\caption{\small The NQR spin-lattice relaxation rate plotted versus temperature normalized
by $T_c$. Calculations are for $2\Delta_0 = 5 \, k_B T_c$ and three
values of the impurity scattering rate $\Gamma$ for unitary scattering. 
Inset: The normalized quasiparticle DOS for corresponding values of 
$\Gamma/\Delta_0 = 0, 0.032, 0.064$.}
\label{fig1}
\end{figure}

\begin{figure}
\centerline{ \includegraphics[height=3.5in]{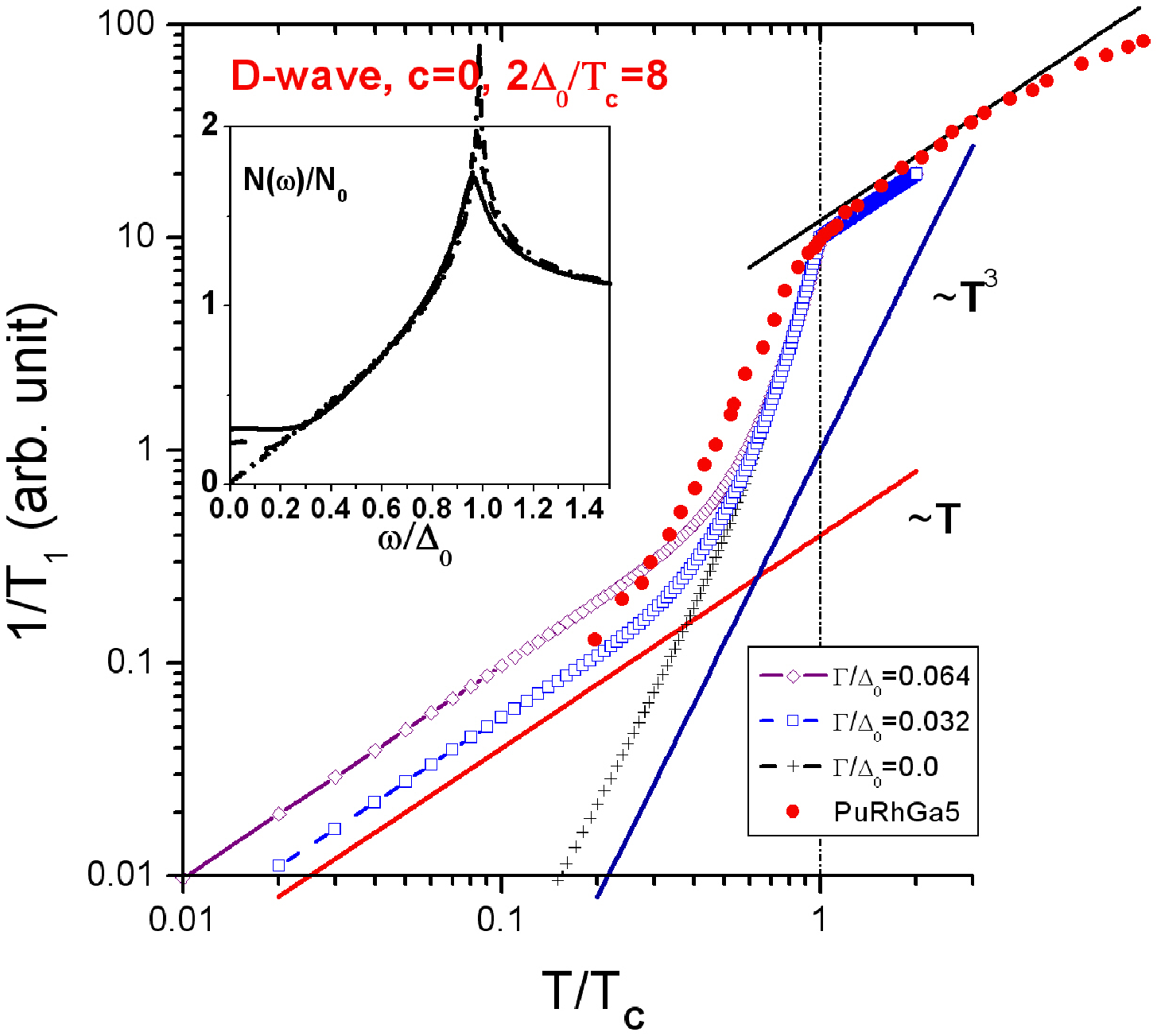} }
\caption{\small The NQR spin-lattice relaxation rate plotted versus temperature normalized
by $T_c$. Calculations are for $2\Delta_0 = 8 \, k_B T_c$ and three
values of the impurity scattering rate $\Gamma$ for unitary scattering. 
Inset: The normalized quasiparticle DOS for corresponding values of
$\Gamma/\Delta_0 = 0, 0.032, 0.064$.}
\label{fig2}
\end{figure}

In figures \ref{fig1} and \ref{fig2}
the spin-lattice relaxation rate of \PRG\ by Sakai et al.~\cite{Sakai}
is shown, where $1/T_1$ is normalized to $1/T_1=10$ at $T=T_{c}$ for ease of comparison.
The insets show the corresponding normalized quasiparticle DOS for varying scattering rates $\Gamma$. 
With our earlier described choice of parameters the impurity scattering rate $\Gamma/\Delta_0=0.032$ is
enough to completely fill the low energy gap with impurity states
and $N(\omega=0)$ reaches more than 25\% of the normal-state DOS $N_0$.

In Fig.~\ref{fig1} we obtain a better fit to the experimental data (symbols) 
assuming a slightly lower superconducting transition temperature 
$T_c = 7.6$ K than their reported value of $T_c = 8.5$ K. This could indicate the presence of a 
pseudogap similar to the high-temperature superconductor YBa$_2$Cu$_3$O$_{7-\delta}$,
where $1/T_1$ is suppressed just above $T_c$.
\noindent
\begin{figure}
\noindent
\centerline{ \includegraphics[height=4.0in]{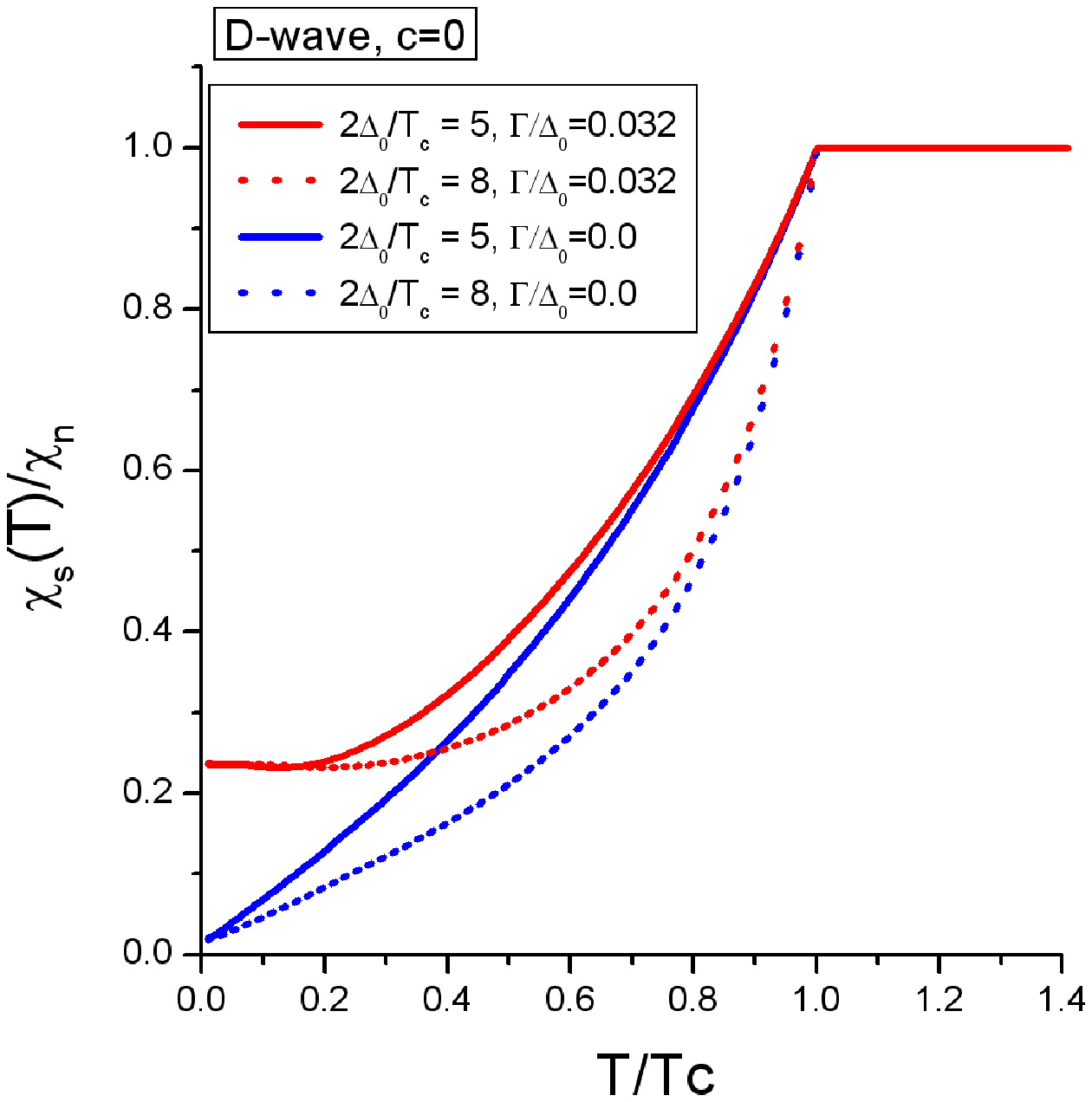} }
\caption{\small The calculated spin susceptibility $\chi_S$ of a d-wave SC normalized by its
normal state value $\chi_{N}$
for gap values $2\Delta_0 = 5 \,{k_B T_c}$ (solid lines) and 
$8 \,{k_B T_c}$ (dotted lines), 
and impurity scattering $\Gamma/\Delta_0 = 0$ and $0.032$. }
\label{fig3}
\end{figure}

For the temperature dependence of the gap,  we chose the parameters
$\beta=1.74$ and $2\Delta_0=5 \,{k_B T_c}$  for the d-wave gap to account for
the strong-coupling effects of superconductivity as explained before. As
expected from the DOS results, due to the impurity induced residual states,
$1/T_1$ displays the linear-$T$ dependence at low temperatures and the region of
$T$-linear behavior increases with  impurity concentration. For a higher value of
$\Gamma/\Delta_0=0.064$, this $T$-linear region extends up to $\sim 0.35$
$T_{c}$. At temperatures near $T_{c}$ the coherence peak is almost invisible
because of the sign-changing gap function, i.e., vanishing of the second term in 
Eq.~(\ref{eq_spin_lattice_rate}). Below $T_{c}$ it shows a nearly $T^3$ behavior due to the
lines of nodes in the gap until it goes
through a gradual crossover region and finally to the $T$-linear region. The
comparison with the experimental data by Sakai et al. \cite{Sakai} on \PRG\
is in good agreement with unitary scattering, a gap value $2\Delta_0 = 5\,{k_B T_c}$,
and a scattering rate close to $\Gamma/\Delta_0=0.032$.
Based on this value, we estimate the superconducting transition temperature of the pristine
sample to be $T_{c0} = T_c + \frac{\pi}{4} \Gamma \approx 8.1$ K or $9.0$ K, depending on
the value of $T_c = 7.6$ K or $8.5$ K.

In Fig.~\ref{fig2}, the normalized $1/T_1$ is plotted for an enhanced strong-coupling 
d-wave gap value $2\Delta_0=8\,{k_B T_c}$, as was recently found for \PCG\ \cite{Curro}.
Due to the larger gap value, the calculated $1/T_1$ is always less than the measured spin-lattice
relaxation rate. Hence we find a poorer fit to the experimental data for this choice of
the strong-coupling gap. 

Fig.~\ref{fig3} shows the prediction for the spin susceptibility, $\chi_S$,  or its
corresponding NMR Knight shift, $K = K_0 + A \chi_S$, where $K_0$ and $A$ are constants for
most materials.
$\chi_S$ is calculated for the same d-wave gap values as was used for the spin-lattice 
relaxation rates in figures \ref{fig1} and \ref{fig2}. Again a modest impurity scattering
rate of $\Gamma/\Delta_0 = 0.032$ results in a large residual susceptibility at zero temperature,
equivalent to roughly $25 \%$ of the normal state DOS or spin susceptibility $\chi_N$.
The quantitative difference in the spin susceptibility between gap values
$2\Delta_0 = 5 \, {k_B T_c}$ and $8\,{k_B T_c}$ should be easily discernible in measurements of 
the Knight shift.

\section{Conclusions}

The NQR spin-lattice relaxation rate $1/T_1$ in \PRG\ is consistent with
a strong-coupling d-wave gap function and unitary impurity scattering similar to the observed
behavior in its sister compound \PCG. 
Inspite of many similarities between \PCG\ and \PRG, we also find marked differences:
(1) The maximum superconducting gap value of \PRG\ is smaller than for \PCG,
i.e., it is $2\Delta_0/{k_B T_c} = 5$ for \PRG\ versus $8$ for \PCG.
This suggests that the fluctuations of the pairing bosons are weaker for \PRG, possibly hinting
at a progressive trend for the strength of the spin-fluctuations in this class of materials. 
(2) Although the \PRG\ sample was of similar age as \PCG\ when measured, 
it had a three times larger relative scattering rate
$\Gamma/\Delta_0 = 0.032$ compared to \PCG\ with $\Gamma/\Delta_0 = 0.01$, 
which could be due to variations of the isotope mix of plutonium between both samples.
(3) $1/T_1$ exhibits a rounded behavior between $T = 7.6$ K and $9$ K resembling the
pseudogap phenomenon in YBa$_2$Cu$_3$O$_{7-\delta}$,
while no such rounding is observed for \PCG.
This certainly needs clarifications by further studies of $1/T_1$ in the normal state,
which will be reported in a separate paper.

\section{Acknowledgments}

We wish to thank John Sarrao,  Joe Thompson and Eric Bauer for many stimulating discussions.
This research was supported by the U.S. Department of Energy at Los Alamos National Laboratory
under contract No. W-7405-ENG-36.  Y. Bang is supported by KOSEF through CSCMR.


\section{References}
\begin{thebibliography}{10}

\bibitem{Sarrao}
J. L. Sarrao, L. A. Morales, J. D. Thompson, B. L. Scott, G. R. Stewart, F. Wastin,
J. Rebizant, P. Boulet, E. Colineau, and G. H. Lander, Nature (London) {\bf 420}, 297 (2002).

\bibitem{Wastin}
F. Wastin, P. Boulet, J. Rebizant, E. Collineau, and G. H. Lander, J. Phys.: Condens. Matter
{\bf 15}, S2279 (2003).

\bibitem{Bang04a}
Y. Bang, A. V. Balatsky, F. Wastin, and J. D. Thompson, Phys. Rev. B {\bf 70}, 104512 (2004).

\bibitem{Curro} 
N. J. Curro, T. Caldwell, E. D. Bauer, L. A. Morales, M. J. Graf, Y. Bang, A. V. Balatsky,
J. D. Thompson, and J. L. Sarrao, Nature (London) {\bf 434}, 622 (2005).

\bibitem{Bauer04}
E. D. Bauer, J. D. Thompson, J. L. Sarrao, L. A. Morales, F. Wastin, J. Rebizant, J. C. Griveau,
P. Javorsky, P. Boulet, E. Colineau, G. H. Lander, and G. R. Stewart, Phys. Rev. Lett. {\bf 93},
147005 (2004).

\bibitem{Sakai} 
H. Sakai, Y. Tokunaga, T. Fujimoto, S. Kambe, R. E. Walstedt, H. Yasuoka, D. Aoki,
Y. Homma, E. Yamamoto, A. Nakamura, Y. Shiokawa, K. Nakajima, Y. Arai, T. D. Matsuda,
Y. Haga, and Y. \={O}nuki, J. Phys. Soc. Jpn. {\bf 74}, 1710 (2005).

\bibitem{Bang03}
Y. Bang, M. J. Graf, and A. V. Balatsky, Phys. Rev. B {\bf 68}, 212504 (2003).

\bibitem{Bang04}
Y. Bang, M. J. Graf, A. V. Balatsky, and J. D. Thompson, Phys. Rev. B {\bf 69}, 014505 (2004).

\bibitem{Hirschfeld86}

P. J. Hirschfeld, D. Vollhardt, and P. W\"olfle, Solid State Commun. {\bf 59}, 111 (1986).

\bibitem{SchmittRink86}
S. Schmitt-Rink, K. Miyake, and C. M. Varma, Phys. Rev. Lett. {\bf 57}, 2575 (1986).

\bibitem{Monien87}
H. Monien, K. Scharnberg, and D. Walker, Solid State Commun. {\bf 63}, 263 (1987).

\bibitem{Bang}
Y. Bang, I. Martin, and A.V. Balatsky, Phys. Rev. B {\bf 66}, 224501
(2002).

\bibitem{Hirschfeld88}
P. J. Hirschfeld, P. W\"olfle, and D. Einzel, Phys. Rev. B {\bf 37}, 83 (1988).

\end{thebibliography}
\end{document}